# Optimization of the power broadening in optically detected magnetic resonance of defect spins in silicon carbide

Jun-Feng Wang[1, 2], Jin-Ming Cui[1,2], Fei-Fei Yan[1, 2], Qiang Li[1, 2], Ze-Di Cheng[1, 2], Zheng-Hao Liu[1, 2], Zhi-Hai Lin[1, 2], Jin-Shi Xu[1, 2,\*], Chuan-Feng Li[1, 2,\*] and Guang-Can Guo[1,2]

[1]*CAS Key Laboratory of Quantum Information, University of Science and Technology of China, Hefei, Anhui 230026, People's Republic of China*
[2] *CAS Center for Excellence in Quantum Information and Quantum Physics, University of Science and Technology of China, Hefei, Anhui 230026, People's Republic of China*
[\*]Corresponding author: jsxu@ustc.edu.cn, cfli@ustc.edu.cn

**Abstract**

Defect spins in silicon carbide have become promising platforms with respect to quantum information processing and quantum sensing. Indeed, the optically detected magnetic resonance (ODMR) of defect spins is the cornerstone of the applications. In this work, we systematically investigate the contrast and linewidth of laser-and microwave power-dependent ODMR with respect to ensemble-divacancy spins in silicon carbide at room temperature. The results suggest that magnetic field sensing sensitivity can be improved by a factor of 10 for the optimized laser and microwave power range. The experiment will be useful for the applications of silicon carbide defects in quantum information processing and ODMR-dependent quantum sensing.

## I. INTRODUCTION

Defects in silicon carbide (SiC) are attractive solid-state quantum systems with respect to quantum photonics, quantum information, and quantum sensing[1–13]. Since

SiC is widely used, its defects with respect to quantum technology are currently drawing a lot more attention[1–13]. Accordingly, multiple bright single-photon emitters with visible and infrared (even telecom) wavelengths in SiC have been investigated for quantum photonics[6–10]. In addition, many researchers have proven that three types of defects (silicon vacancy, divacancy, and nitrogen vacancy center) spins can be controlled using laser and microwave (MW); moreover, they have shown great promise with respect to quantum information processing and quantum sensing[1–5,11–17]. In particular, due to their near-telecom fluorescence spectrum[1–3,16] and long coherence time even at room temperature[1–3,16,18], divacancy defects in SiC have realized high-fidelity quantum register[11], spin-photon interface[16], and quantum sensing, the latter of which includes magnetic field sensing[1,2], electric field sensing[12], broad-range temperature sensing[14,15], and strain sensing[13].

For the majority of applications, the continuous-wave optically detected magnetic resonance (ODMR) of SiC defects is vital. In particular, the optimized ODMR signal is critical for high quantum-sensing sensitivity. For example, although we can detect temperature by using sensing-pulse sequencing[14,15,19], temperature-dependent ODMR is used because it is a more direct method[14,15,20–22]. Moreover, although Ramsey methods can realize high sensitivity for direct current (DC) magnetic field sensing[23], the simplest method is to directly measure the Zeeman splitting in the ODMR signal[24–29]. In addition, for the DC strain and electric field sensing, the optimized ODMR signal is important with respect to achieving high sensitivity[12,13]. Indeed, for the electron-nuclear spin quantum register, it is crucial to decrease the ODMR linewidth in order to identify strongly coupled electron-nuclear hybrid pairs[3,11,16]. Moreover, since the ODMR contrast of SiC defects is only roughly 1%, it is important to optimize ODMR contrast in order to conveniently implement the experiments[1–5]. In light of those information, it is vital to optimize the ODMR signal of SiC defect spins, which have not been reported before.

In this work, we study the ODMR contrast and linewidth as a function of laser and MW power with respect to the PL6 divacancy-defect ensemble in 4H-SiC at room temperature. First, the coherent control of the PL6 defect spins in 4H-SiC is conducted at room temperature. Then, we study the laser- and MW-dependent ODMR contrast and linewidth of PL6 defects in 4H-SiC. Finally, we accurately measure the magnetic field by optimizing the laser and MW power. Indeed, the experiment will be useful for ODMR-dependent quantum sensing, including magnetic field and temperature sensing using technology-friendly SiC materials.

## II. EXPERIMENT AND RESULTS

A commercially available bulk high-quality semi-insulating 4H-SiC (Cree) was used in the experiment[14,15]. A homebuilt room-temperature confocal system combined with a MW system was used to polarize and control the divacancy-defect spin[14,15]. A 920 nm laser was used to excite the sample through a 1.3 NA oil objective (Nikon). The divacancy fluorescence (above 1,000 nm) was collected by a photoreceiver (Femto, OE-200-IN1, 1.1 kHz time constants) through a multimode fiber[1,14]. A 50 μm diameter copper wire was used to radiate MW in order to control the defect spins. During the experiment, we used a lock-in method (20-Hz modulated, SR830 Lock-In Amplifier) to read out the signal[1,14,30] and an electromagnet to generate an c-axis magnetic field (perpendicular to the sample surface).

The neutral divacancy defects in 4H-SiC are composed of adjacent carbon and silicon vacancies. There are seven types (PL1-7) of divacancy defects in 4H-SiC. Among these defects, PL1, (2, 6) are c-axis divacancy defects while the PL3, (4, 5, 7) are basal divacancy defects. In this work, we focus on the c-axis PL6 defects, which are identified by the magnetic field-dependent ODMR signal. The defects exhibit spin-1 ground states, which can be controlled at room temperature. The ground-state spin Hamiltonian of the divacancy defects is

$$H = DS_Z^2 + E(S_X^2 - S_Y^2) + g\mu_B \mathbf{B} \cdot \mathbf{S},  \quad (1)$$

where $D$ and $E$ denote the axially symmetric (Z) and anisotropic (X and Y) components of the zero-field-splitting (ZFS) parameter, respectively, resulting in two zero-field resonant frequencies defined by $(D \pm E)$[1,2,14]. $g = 2$ is the electron g-factor,

$\mu_B$ denotes the Bohr magneton, **B** denotes the applied axial static magnetic field, **S** with the three components Sz, Sx and Sy denotes the vector of electronic S = 1 spin matrices. When the spin in the ground state is excited by a resonant MW field, there will be a change in the fluorescence emission intensity (ODMR).

Fig. 1(a) shows the photoluminescence (PL) intensity as a function of laser power (*P*). Herein, the red line is the fit of the data using the power-dependence model, *I(P) = $I_s$/(1 + $P_0$/P)*, where $I_s$ is the maximal count and $P_0$ is the saturation power. Indeed, the fit indicates that the maximal intensity is 7.9 ± 0.3 V and the saturation power was 29.4 ± 0.5 mW/μm². In order to turn the voltage detected by the Femto photoreceiver to photon counts, we compare the counts of the superconducting single-photon detector (Scontel)[8,15] with the voltage of the Femto (Fig. 1(b)). Inferred from the fitting, the conversation of 1V voltage to photon counts is 14.0 ± 0.1 Mcps[14] (million counts per second). The background noise of the Femto is measured to be 0.8 mV, which is much smaller than the detected results and is omitted. In the experiment, we used an oil objective with a high NA (1.3). The confocal spot size of the laser beam was roughly 1 μm² in the x–y plane and 1 μm in the z-direction. Combining the detected Femto voltage[14] and the count of a single divacancy defect[16], the divacancy density in our sample was (1.00 ± 0.02) × 10³ μm⁻³, which is similar to the previous results[14]. The MW scanning at zero magnetic field is shown in Fig. 1(c). ΔPL represents the change of the photoluminescence intensity and PL represents the intensity with the off-resonant MW field. Several resonant frequencies are observed, which correspond to the zero-field splitting (ZFS) parameters of PL5 (1,343.5 and 1,374.9 MHz), PL6 (1,350 and 1,352.5 MHz) and PL7 (1,333.2 MHz), respectively. Accordingly, the *D* and *E* values of the ZFS of PL6 defects are 1,351.2 MHz and 1.3 MHz, respectively. The ODMR signals of the divacancy as a function of the c-axial magnetic field are shown in Fig. 1(d). The two transitions of PL6 defect spins split at a slope of 2.8 MHz/G under the c-axis magnetic field.

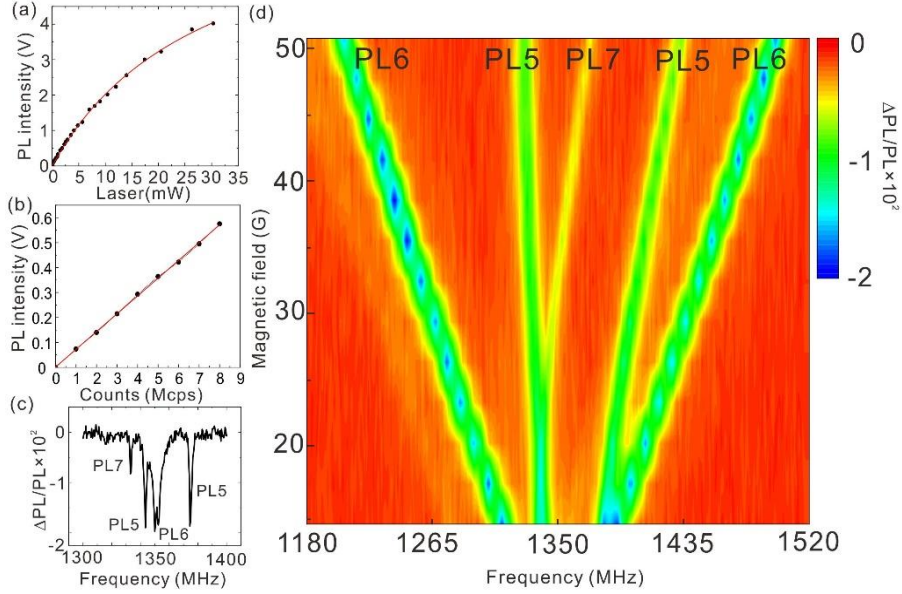

**FIG. 1.** Saturation curve and ODMR spectrum of divacancy defects. (a) The saturation behavior of divacancy defects at different laser powers. (b) Comparison of the counts of the superconducting single-photon detector and the voltage of the Femto photoreceiver. (c) The ODMR spectrum of divacancy defects at a magnetic field of zero. (d) The ODMR spectrum of divacancy defects as a function of the magnetic field.

We coherently control the PL6 defect spins with a magnetic field of 32.5 G at room temperature. In the experiment, the left branch was selected at a transition frequency of 1,260 MHz. In the experiment of coherent control of defect spins, two laser pulses with the duration of 5 μs are used to polarize and read the spin states, respectively. To measure the Rabi oscillation, a microwave pulse at the resonant frequency with variable lengths is used to control defect spins. Two π/2 microwave pulses implemented at different time are used for the Ramsey oscillation experiment. Fig. 2(a) shows three Rabi oscillations with different MW powers. The Rabi frequencies linearly increase as a function of the square root of the MW power, as shown in Fig. 2(b). Therefore, we can use the Rabi frequency to represent the MW power. Ramsey oscillation reflecting the free induction decay time $T_2^*$, which is shown in Fig. 2(c). The 4.4 MHz oscillation was due to MW detuning. From the data fit, we can infer that the dephasing time ($T_2^*$) was 0.5 ± 0.1 μs. Moreover, in order to observe the electron spin-echo envelope modulation, a standard spin echo π/2-π-π/2 microwave pulses were used for coherent time measurement[1]. We measured the spin echo at a strong magnetic field of 190 G. The results are shown in Fig. 2(d), from which it is evident that the coherent time $T_2$ was 33.6 ± 2.0 μs, which, alongside $T_2^*$, are consistent with the previous results[1].

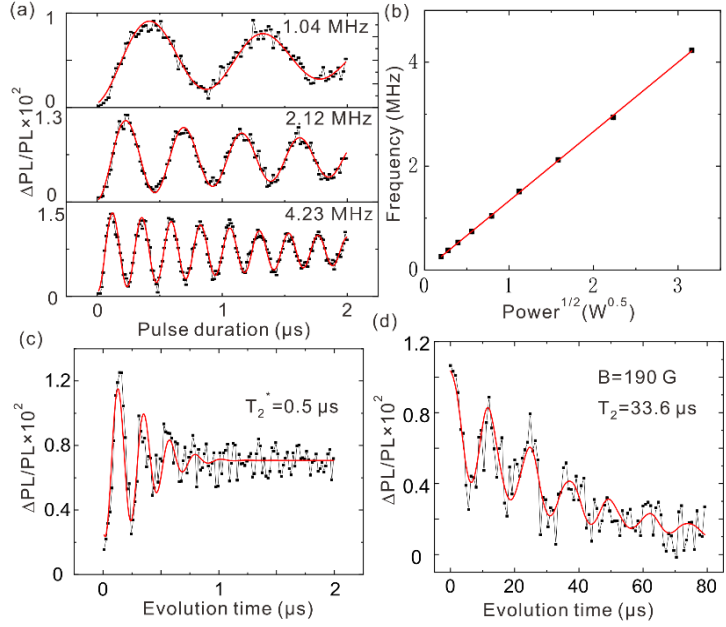

**FIG. 2.** Coherent control of PL6 defects at room temperature. (a) Three Rabi oscillations with different MW powers (the red lines represent data fitting using the exponentially decaying cosine function, and the inset shows the Rabi frequencies). (b) Rabi frequencies increase linearly with the square root of the total MW power on the copper wire (the red line is a linear fit). (c) Ramsey oscillaltion of the defects (the MW frequency detunes for 4.4 ± 0.2 MHz). (d) Spin-echo measurement of the defect spins (the red line represents data fitting using the decaying exponential function).

Following this, we studied the laser- and MW-dependent ODMR signals of the PL6 defects. The linewidth of the ODMR spectrum was fundamentally limited by the inhomogeneous dephasing time $T_2^*$ of the PL6 spin defects[24,25]. $T_2^*$ was determined by the magnetic dipolar interactions between the PL6 electron spin and the impurities spin bath (including the $^{29}$Si (4.7%, $I_{Si}$ = 1/2) and $^{13}$C (1.1%, $I_C$ = 1/2) nuclear spins)[18,31]. Moreover, the laser- and MW-dependent power broadening also had an effect on the ODMR linewidth. In order to achieve high sensitivity with respect to ODMR-dependent quantum sensing, the ODMR signal should be narrow and of a high contrast. Accordingly, we studied ODMR contrast and linewidth as functions of laser power and Rabi frequency representing the MW power. Fig. 3(a) shows the ODMR measurement of divacancy with a laser density of 4 mW/μm$^2$ and a Rabi frequency of 2.1 MHz at the magnetic field of 32.5 G. We selected the left branch ($|0\rangle \leftrightarrow |-1\rangle$) transition of the PL6 defects. Moreover, inferring from the Lorentzian function fit, the resonant frequency was 1,260 ± 1 MHz with a contrast of 1.8 ± 0.1 %; the full width at half maximum (FWHM) was 11.9 ± 0.2 MHz. The ODMR signal of the PL6 left branch with low laser (1 mW/μm$^2$) and MW power (0.18 MHz Rabi frequency) is shown in Fig. 3(b).

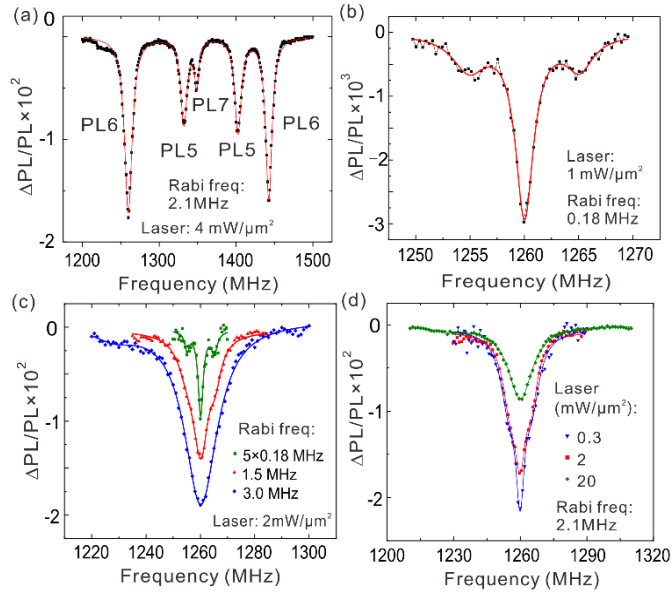

**FIG. 3.** ODMR signals of PL6 defects. (a) The ODMR measurement of defects with a wide MW frequency scanning range at a magnetic field of 32.5 G (the red line represents the data fit using the Lorentzian function). (b) The ODMR measurement of the left branch of the PL6 at low laser power (1 mW/μm$^2$) and 0.18 MHz Rabi frequency (the red line is the data fit using three Lorentzian functions, the center peak consists of the PL6 electron spin signals that were not strongly coupled to any nuclear spins, and the two small peaks correspond to the PL6 electron spins signals that were strongly coupled with $^{29}$Si nuclear spins of 9.8 ± 0.2 MHz). (c) Three representative ODMR measurements at different Rabi frequencies for a laser power of 2 mW/μm$^2$ (the ODMR signal was magnified five times for 0.18 MHz MW power). (d) Three representative ODMR measurements at different laser powers for a Rabi frequency of 2.1 MHz (a sum of three Lorentzian peaks was used to fit the data at a laser power of 0.3 mW/μm$^2$ and 2 mW/μm$^2$).

We can see three obvious ODMR peaks; the strong center peaks correspond to the PL6 electron spin signals that were not strongly coupled to any nuclear spins. Inferring from the fit, the contrast was 0.29 % and the linewidth was 2.02 ± 0.06 MHz. The pronounced doublet was due to the strong hyperfine coupling (9.8 ± 0.2 MHz) between some PL6 electron spins and $^{29}$Si nuclear spins, which is similar to the previous results[11]. Therefore, if we want to observe strongly coupled electron-nuclear spin pairs, it is important to decrease the ODMR linewidth[11]. Fig. 3(c) shows three ODMR signals at different MW powers for a laser power of 2 mW/μm$^2$. Obviously, both the ODMR contrast and linewidth increase with an increase in MW power. Specifically, the ODMR contrast (and linewidth) for Rabi frequencies of 0.18 MHz and 3 MHz are 0.22 % (2.3 MHz) and 1.88 % (14.7 MHz), respectively. Three ODMR signals at different laser powers for the same Rabi frequency (2.1 MHz) are shown in Fig. 3(d). The ODMR contrast obviously decreases as the laser power decreases, while the ODMR linewidth increases as the laser power increases. We can see that the ODMR signals have obvious hyperfine coupled $^{29}$Si-induced doublets besides the strong center peak under a laser power of 0.3 mW/μm$^2$ and 2 mW/μm$^2$. For a high laser power of 20 mW/μm$^2$, there is only one strong ODMR peak which is the result of the superimposition of the three ODMR peaks. Indeed, the results demonstrate that laser and MW power result in the power broadening of the ODMR signal.

We systematically studied the contrast and linewidth of laser- and MW power-dependent ODMR. Fig. 4(a) summarizes the dependence of ODMR contrast with laser power. For all three different MW powers, the contrast decreases as the laser power increases. In particular, for a Rabi frequency of 2.94 MHz, the ODMR contrast decreases from 2.19 % to 1.05 % as the laser power increases from 0.3 mW/μm$^2$ to 4 mW/μm$^2$. The ODMR contrast slightly decreases when the laser power increases from 0.3 mW/μm$^2$ to 4 mW/μm$^2$, while, when the laser increases from 4 mW/μm$^2$ to 20 mW/μm$^2$, the contrast decreases rapidly. The ODMR linewidth (measured as FWHM), as a function of laser power, at three different MW powers is presented in Fig. 4(b). The ODMR linewidth only slightly increases as the laser power increases, which is significantly smaller than that of the nitrogen vacancy center in diamond[24]. The ODMR contrast, as a function of Rabi frequency, with three different laser powers is shown in Fig. 4(c), from which it is evident that contrast increases as the Rabi frequency increases and that the maximum has a saturation value of 2.3 ± 0.1 %. Fig. 4(d) shows the ODMR linewidth at three different laser powers as a function of Rabi frequency. The linewidth slowly increases when the Rabi frequency is below 1 MHz. However, the linewidth rapidly increases when the Rabi frequency is larger than 1 MHz. This is due to the superimposition of the center peaks and the hyperfine coupled $^{29}$Si-induced doublets.

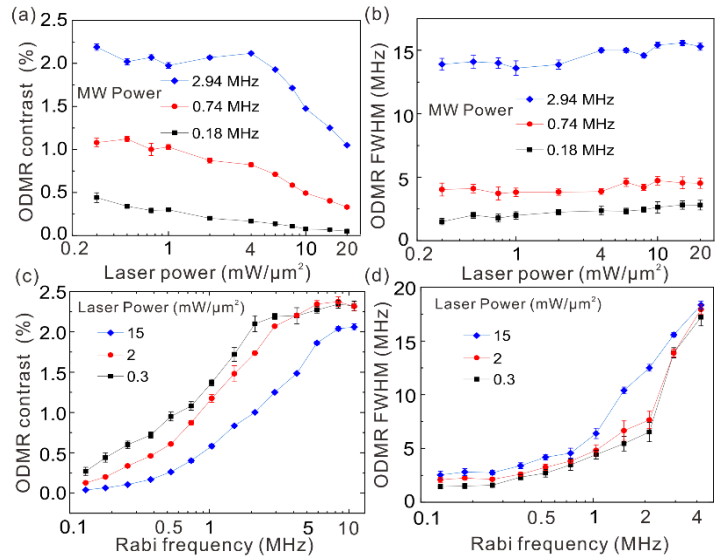

**FIG. 4.** Laser- and MW power-dependent ODMR contrast and linewidth. (a) and (b) represent the ODMR contrast and linewidth of three different Rabi frequencies as a function of laser power plotted in log scale, respectively. (c) and (d) represent the ODMR contrast and linewidth of three different laser powers as a function of Rabi frequency plotted in log scale, respectively.

The optimized ODMR is important for SiC-based quantum sensing, including DC magnetic field sensing[23–30] and broad-range temperature sensing[14,15,20–22], etc. Therefore, we summarized the DC magnetic field sensitivities using the experimental ODMR signals at different laser and MW powers. The shot-noise-limited relative magnetic field sensitivity is $\eta_B \approx \frac{h}{g\mu_B} \frac{\Delta\nu}{C\sqrt{R}}$, where $h$ is the Planck constant, $R$ is the rate of detected photons, $C$ is the contrast of the ODMR, and $\Delta\nu$ is the ODMR width [23–29]. Fig. 5 shows the contour plot of sensitivity as a function of laser and MW power. In the experiment, we used 121 (11 × 11) different laser and MW power settings in order to generate the contour plot. The best relative sensitivity, $\eta_B$=(4.0 ± 0.2) $\mu T/\sqrt{Hz}$, was reached in the optimized range of the laser from 2 mW/μm² to 8 mW/μm²; for MW power, it was from 0.75 MHz to 1.75 MHz. Indeed, the sensitivity can be improved by a factor of 10 for the optimized laser and MW power range. Moreover, ODMR-dependent temperature sensing has a similar optimized laser and MW power range compared to that of magnetic field sensing[14,15,21–22].

As a preliminary experiment of direct magnetic field sensing, we used a Helmholtz coil to generate an additional c-axis magnetic field. First, we measured the ODMR as a function of a larger electric current in order to calibrate the magnetic field generated by the Helmholtz coil. In the experiment, we used 4 mW/μm² laser pumping and a Rabi frequency of 1.51 MHz MW to measure the ODMR signals, thereby obtaining the optimized magnetic sensitivity. Two ODMR measurements (without (black) and with the magnetic field (red, 1.74 G)) are shown in Fig. 5(b). Fig. 5(c) shows the ODMR resonant frequency as a function of large magnetic fields (Helmholtz coil). The slope is 2.83 ± 0.02 MHz/G, which suggests that the magnetic field is along the c-axis. Following this, we measured small magnetic fields using the ODMR spectrum of the divacancy defects. Fig. 5(d) shows the two ODMR spectra, without and with the little magnetic field. There are 100 data points with the integration time per point being 4 seconds. The inset is a zoom-in of the ODMR peaks. Inferring from the fit, the resonant frequencies are 1,259.87 ± 0.04 MHz and 1,259.68 ± 0.04 MHz, respectively. The corresponding magnetic field is 6.8 ± 1.4 µT, which demonstrates a high precise direct magnetic field sensing. Fig. 5(e) shows the ODMR resonant frequency as a function of smaller magnetic fields.

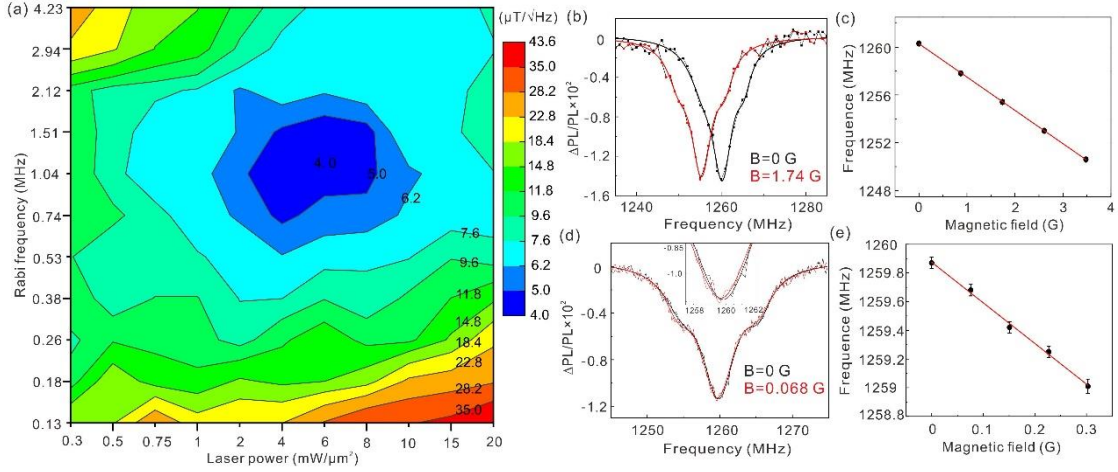

**FIG. 5.** (a) Two-dimensional plot of the experimental magnetic field sensitivity as a function of laser power and Rabi frequency (the lines represent the isomagnetic field sensitivities). (b) and (d) consist of ODMR measurements without and with a magnetic field (generated by the Helmholtz coil), respectively. (c) and (e) consist of ODMR resonant frequencies as a function of strong and weak magnetic fields (generated by the Helmholtz coil), respectively (the red lines represent data fittings).

Futhermore, the magnetic field can also be measured by Ramsey method, and the corresponding magnetic field sensitivity is calculated as $\eta_B \approx \frac{\hbar}{g\mu_B} \frac{1}{C\sqrt{RT_L}} \times \frac{1}{\sqrt{T_2^*}}$, where $T_2^*$ is 0.5 µs and the laser readout time ($T_L$) is 300 ns. Under 4 mW/µm² laser pumping and a Rabi frequency of 1.5 MHz the corresponding counts (*R*) and contrast (*C*) are 14 Mcps and 1.45%, respectively. The sensitivity is (280 ± 15) $nT/\sqrt{Hz}$ which is almost 14 times smaller than the optimized ODMR sensitivity. Indeed, combined with a high concentration of defects[2,30] as well as efficient detection methods, including solid immersion lenses[5], nanopillars[7], and photonic crystal cavities[8,9], the magnetic field sensitivity can increase significantly.

### III. CONCLUSIONS

In this work, we investigated the laser and MW power-broadening ODMR spectra of PL6 divacancy-defect spins at room temperature under a small c-axis magnetic field. The results suggest that both ODMR linewidth and contrast increase as the MW power increases. Moreover, the ODMR linewidth also increases as the laser power increases, while the ODMR contrast decreases as the laser power decreases. Based on this, we also summarized the DC magnetic field sensitivities using the experimental ODMR signals with different laser and MW powers, improving the magnetic fields sensing sensitivity by a factor of 10 for the optimized laser and microwave power range. Indeed, the results pave the way for the application of technology-friendly SiC defect spins using ODMR signals with respect to quantum information processing and high sensitivity quantum sensing.

**ACKNOWLEDGMENTS**

J. F. Wang and J. M. Cui contributed equally to this work. This work was supported by the National Key Research and Development Program of China (Grant No. 2016YFA0302700), the National Natural Science Foundation of China (Grants No. 61725504, 61327901, 61490711, 61905233, 11975221,11821404 and 11774335), the Key Research Program of Frontier Sciences, Chinese Academy of Sciences (CAS) (Grant No. QYZDY-SSW-SLH003), Anhui Initiative in Quantum Information Technologies (AHY060300 and AHY020100), the Fundamental Research Funds for the Central Universities (Grant No. WK2030380017 and WK2470000026). This work was partially carried out at the USTC Center for Micro and Nanoscale Research and Fabrication.